\documentclass[11pt]{article}
\usepackage{aas2pp4}
\def\kmsmpc{${\rm km\,s^{-1}\,Mpc^{-1}}$} 

\def\Ho{${\rm H_{o}}$}                          
\begin{document}
\slugcomment{To Appear in {\bf {\em Nature}}, December 10, 1998}
\title {\large {\bf Galaxy Disruption as the Origin of Intracluster Light in
the Coma Cluster}}

\bigskip

\centerline{{Michael~D.~Gregg}\footnote{\it Physics Department,
University of California, Davis, California 95616, and Institute for
Geophysics and Planetary Physics, Lawrence Livermore National
Laboratory, L-413 Livermore, California 94550, USA} \&
{Michael~J.~West}\footnote{\it Department of Astronomy and Physics,
Saint Mary's University, Halifax, Nova Scotia B3H 3C3, Canada}}

\bigskip

{\bf 

Although the existence of a faint background of starlight in the core
of the Coma cluster has been well established$^{1,2}$, its origin is
uncertain.  This vast sea of stars could have formed outside the
galaxies, using gas left over from the time of the cluster's birth.
Alternatively, it might be the accumulated debris generated by
interactions between the galaxies over the lifetime of the
cluster$^{3}$.  Here we report the discovery of three large, low
surface brightness features in the Coma cluster; the most spectacular
is a plume-like structure, 130 kpc long, in the cluster's heart.
These structures will disperse over the next one to two billion years,
thereby enhancing the general background light.  If this epoch is
typical, we argue that a significant fraction -- perhaps even most --
of the intracluster light results from a steady accumulation of tidal
debris generated during galaxy-galaxy and galaxy-cluster
interactions.
}

\vspace{1in}

\noindent {To Appear in {\bf {\em Nature}}, December 10, 1998}

\vspace{3in}

\pagebreak

\twocolumn

We have been using the Kitt Peak Burrell Schmidt\footnote{Kitt Peak
National Observatory, National Optical Astronomy Observatory, is
operated by the Association of Universities for Research in Astronomy,
Inc. (AURA), under cooperative agreement with the National Science
Foundation.  Observations made with the Burrell Schmidt telescope of
the Warner and Swasey Observatory, Case Western Reserve University.}
0.6 metre telescope to conduct a multi-colour survey of the Coma
cluster and its environs over a $3^\circ \times 3^\circ$ area.  The
CCD on the Schmidt has $2\arcsec$ pixels and a field of view of
$1.1^\circ$.  In March of 1996, we took a series of images centred
midway between NGC4874 and 4889, the supergiant elliptical galaxies
which dominate the centre of the Coma cluster.


Inspection of the final R-band image (Fig.~1) reveals the presence of
three large, low surface brightness (LSB) features.  Their locations
and approximate sizes are indicated by the heavy solid lines labeled
'1', '2', and '3' in Fig.~1.  All of these objects can be seen in the
B, V, and I frames as well, confirming their reality.  All are much
larger and more prominent than the recently reported slim arc of tidal
debris in the Coma cluster$^{4}$ (no.~4 in Fig.~1).  Given the myriad
images of the Coma cluster that have been obtained over the years, it
is surprising that these features have escaped notice until now.
There are at least two instances in the literature of images that show
one or more of them clearly$^{5,6}$, but without any comment by the
authors.  To isolate the low surface brightness (LSB) features,
we digitally `cleaned' portions of the images to remove superposed
galaxies and stars by interpolation with background replacement
(Fig.~2 and 3).

\noindent {\bf {\em LSB object no.\ 1}}~~~The most striking of the
features is a plume-like structure adjacent to NGC4874 (Figs~1
and~2).  It extends for at least 4.5\arcmin\ and varies from
30\arcsec\ to 60\arcsec\ in width; it may well extend even farther to
the west, past the large elliptical galaxy NGC4864.  Adopting a
distance of 100~Mpc to Coma, the plume is $\sim130$ kpc in
length and varies from 15 to 30 kpc wide.  The integrated
apparent magnitude is $R = 15.6\pm0.1$, though it may contain more
material outside our adopted borders.  The mean R-band surface
brightness is $\mu_R = 25.7$ mag/arcsec$^2$.  The colours of the plume
are consistent with those of normal elliptical galaxies of
intermediate brightness,
although the photometric uncertainties and the removal of superposed
objects leave open the possibility of significant star formation in
the LSB material.

\noindent {\bf {\em LSB object no.\ 2}}~~~This feature is
a pool of diffuse light adjacent to a group of galaxies including
IC3957, IC3959, and IC3963 (upper panel of Fig.~3.)  It is located
near one end of a broader, extended swath of even lower surface
brightness cluster background light$^{5}$, linking it to NGC4874.
This association makes it likely that the feature is within
the Coma cluster core.  The pool of material has a roughly circular
distribution, $\sim 80\arcsec$ in diameter ($\sim 40$~kpc).  Its
colours and surface brightness are similar to those of the plume, with
a total integrated magnitude 0.6 fainter.

\noindent {\bf {\em LSB object no.\ 3}}~~~The third LSB feature is
quite chaotic in appearance (lower panel, Fig.~3).  The lowest
surface brightness portions of this object form a bridge to the
outskirts of the Coma member galaxy NGC4911, where there is additional
low surface brightness material (inset, Fig.~3); the association
places the feature in the Coma cluster.  Several small knots are
visible in the highest surface brightness portion, suggestive of star
formation; however, its colours are not particularly blue as would be
expected in that case.  Its overall photometric properties are similar
to the other LSB features.

\noindent {\bf {\em LSB object no.\ 4}}~~has already been described in
detail by Trentham \& Mobasher$^{4}$ as a ``giant low surface
brightness arc of length $\sim 80$ kpc.''  From our low resolution
Schmidt images, we cannot confidently recognize this feature as a
tidal arc rather than a curious alignment of faint galaxies, nor can
we analyse it in any detail.  Its total length is 180\arcsec\ with a
mean B surface brightness of $< 26.5$ mag arcsec$^{-2}$ (ref.~4); in
our R-band image, its average width is $\sim 5\arcsec$.  Assuming
typical galaxy colours, we estimate the integrated R magnitude as about
18, roughly 5-10 times fainter than the other LSB features reported
here, with a comparable mean surface brightness.


The LSB features all appear to be non-equili\-bri\-um configurations of
stars deep within the gravitational potential well of the Coma
cluster.  We believe that they are best interpreted as transient
features produced either by galaxy-galaxy interactions or by stripping
of galaxies by the global cluster tidal field.  Coma's large
$\sim1,000$~km s$^{-1}$ velocity dispersion would seem to preclude strong
galaxy-galaxy interactions since the energy transferred between
systems is roughly proportional to the time spent in close proximity;
typical high-speed encounters between galaxies will not produce
significant disturbances$^{7}$.  Yet the presence of these LSB objects
shows that Coma's galaxy population is undergoing vigorous dynamical
evolution.  There is now ample evidence from both optical and X-ray
observations that the Coma cluster is a dynamically young system with
several components in the throes of merging$^{2,8-12}$.  The
chaotic nature of subcluster mergers and the relatively low internal
velocity dispersions of constituent subgroups, such as the recently
identified ``galaxy aggregates'' in Coma$^{13}$, provide
opportunities for lower speed encounters, producing both galaxy-galaxy
and galaxy-cluster tidal interactions.

Linear features such as the plume are routinely produced in numerical
simulations of galaxy tidal interactions$^{14,15}$, although the
plume's origin is hard to pinpoint from our data.  It may be the
result of interaction between the two elliptical-like nuclei embedded
near its centre, though their large velocity difference ($> 1,100$
km s${-1}$)$^{16}$ makes it unlikely that they could have generated such a
long tidal feature while remaining apparently close together$^{14}$.
Tides acting in the outer regions of NGC4864 (the large elliptical
galaxy located near the plume's western end) could also produce the
plume.  Another possibility, perhaps less likely, is that the plume is
the remains of a galaxy being completely disrupted by the global tidal
field as it plunges into the cluster core, not unlike the dive into
Jupiter of comet Shoemaker-Levy.

The amorphous structures of the LSB features 2-4 are not typical of
what is seen in simulations, or examples on the sky, of tidal debris;
nevertheless, tidal interaction is a natural explanation.  LSB object
no.~2 may be the product of strong tidal interaction within the
subgroup containing IC3959 and IC3963; their velocity difference is
only 240 km s$^{-1}$ (ref.~16).  NGC4911 and LSB object no.~3 lie within a
$\sim 1$~Mpc-long ridge of heightened X-ray emission (see Fig.~1)
which may have been produced by tidal disruption of a galaxy group on
its recent passage through the cluster core$^{11}$.  This suggests
that the LSB object no.~3 may have originated during the same dynamical
interaction that has brightened the X-ray emission.  It could be the
remains of a galaxy that has completely disrupted or it may be
material that has been stripped from the outskirts of NGC4911.  The
Trentham \& Mobasher object is less clearly associated with cluster
member galaxies, but it too is best explained by tidal
processes$^{4}$.

With four examples of rather severe tidal interactions, one should
expect many more less spectacular encounters to be taking place.  The
coarse resolution Schmidt images, however, may be incapable of
revealing them as such; LSB feature no.~4 is detected, but its nature is
not apparent.  Our data suggest that there are many weaker tidal
interactions and galaxy disturbances taking place in Coma, but finer
images which reach at least as deep will be necessary to analyse them.
If evidence for weaker interactions is found, it will help account for
the dearth of low density dwarf galaxies in Coma's core$^{17}$, as
such galaxies are most susceptible to tidal disruption by the
cluster's strong gravitational field, and also the anti-correlation
between the luminosity of the brightest galaxy in a cluster and the
size of the dwarf galaxy population.$^{18}$.

The LSB features will gradually disperse, either blending into the
existing background of diffuse intracluster material or accreting onto
the giant galaxies NGC4889 and NGC4874 at the bottom of Coma's
gravitational potential well.  All four tidal features are located
within $\sim 0.5$ Mpc of the cluster center and should dissipate in no
more than a few cluster crossing times, about 1 to 2 billion years.
If this material contains quickly evolving massive stars or even
ongoing star formation, the use of brightest cluster galaxies as
standard candles is further complicated.$^{19,20}$
The very highest velocity material will contribute to the population
of intergalactic stars and perhaps intergalactic globular clusters as
well; evidence of such components may have been
detected$^{21,22,23,24}$ in other clusters.  If we are viewing Coma at
a typical moment in its history, then the presence of {\it at least}
four large transient star piles indicates that a substantial amount of
material could have been liberated from galaxies over the cluster's
lifetime; every $\sim 1$~Gyr, the equivalent of a galaxy of R-band
absolute magnitude M$_R \approx -19$ is being disrupted and accreted
by the cluster core.  Integrated over the 10 to 20 billion year
history of the Universe, the amount of stellar material added to the
pool of intracluster light would be equivalent to an object with M$_R
\approx -22$.  This is about 20\% of the total luminosity of either
NGC4874 or NGC4889, and the interaction rate was almost certainly
higher in the past.  A significant portion of the mass of the central
galaxies may consist of material accreted from tidally dismantled
galaxies as the cluster dynamically evolves$^{18,20}$.  The LSB
objects now in Coma provide a vivid 'snapshot' of this process in
action.

\acknowledgments

MJW acknowledges a research grant from NSERC of Canada.  Part of the
work reported here was done at the Institute of Geophysics and
Planetary Physics, under the auspices of the U.S. Department of Energy
by Lawrence Livermore National Laboratory.  We acknowledge use of the
NASA/IPAC Extragalactic Database (NED) which is operated by the Jet
Propulsion Laboratory, Caltech, under contract with the National
Aeronautics and Space Administration.

\pagebreak

\onecolumn

\begin{figure}
\plotfiddle{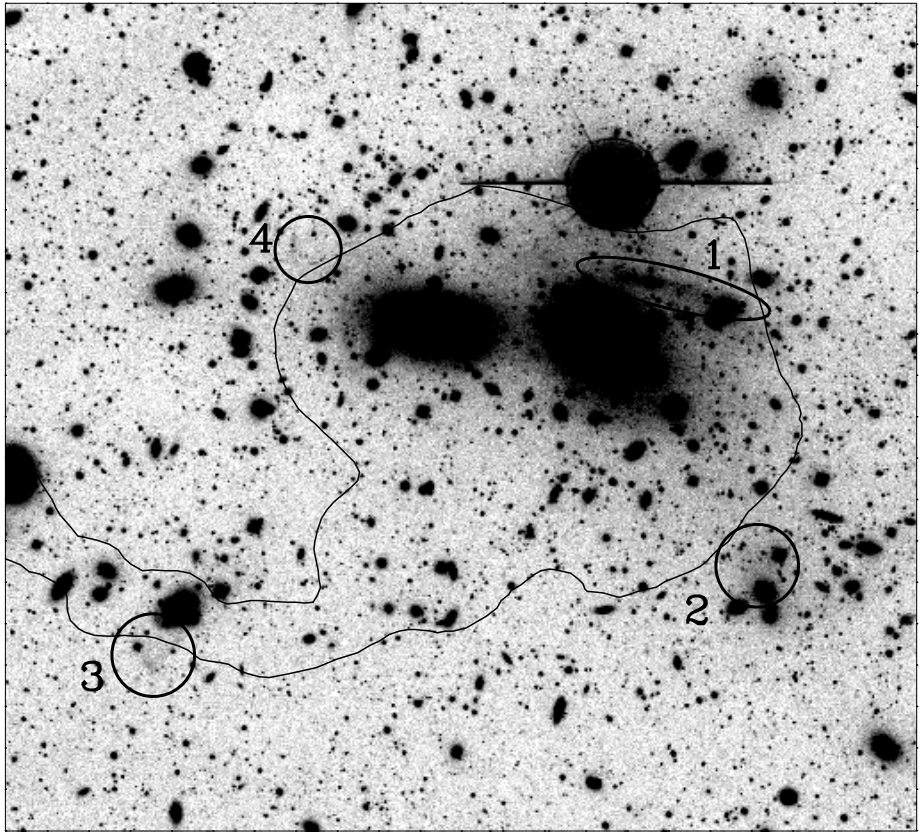}{4.in}{0}{100}{100}{-235}{-35}
\caption{R-band image of the Coma cluster core showing the locations
and approximate sizes of the low surface brightness features (heavy
solid circles labeled 1-4) discussed here.  The image is $37\arcmin
\times 33 \arcmin$; north is up, east to the left.  Observing
conditions were photometric with good seeing, and there was no Moon.
The R-band integration totals 75 min.; the B, V, and I band
images total 30-45 min.  The images were reduced using standard
procedures and combined using image masks to remove cosmetic defects
on the CCD.  Flatfielding using twilight images resulted in
peak-to-peak sky flatness of $\sim 2\%$ in all colours.  The thin solid
line is one X-ray contour from ROSAT observations$^{11}$ showing the
extended ridge which includes LSB object no.~3.}
\end{figure}

\twocolumn

\begin{figure}
\plotfiddle{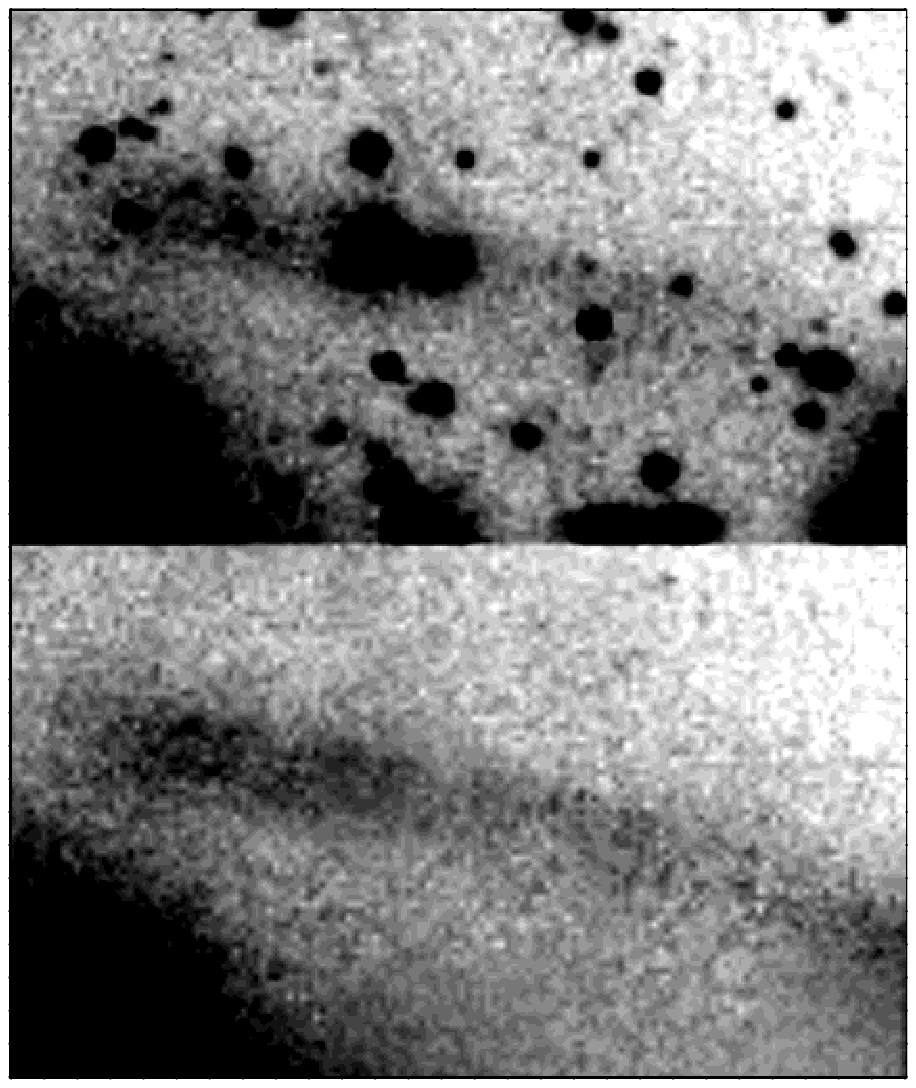}{4.5in}{0}{90}{90}{-170}{-30}
\end{figure}

\begin{figure}
\caption{The plume feature before (top) and after (bottom) removing
foreground objects.  North is up, east is to the left.  The sloping
background from the giant elliptical NGC4874 was fit and removed and
the remaining counts from the plume summed to determine a total
brightness.  The plume is roughly $4\farcm5 \times 1\arcmin$ in
extent, with an integrated R-band magnitude M$_R = 15.6$ and mean R-band surface
brightness of $\sim 26.3$ mag arcsec$^{-2}$.  Adopting a distance
of 100~Mpc for the Coma cluster (equivalent to assuming \Ho $\approx
70$ \kmsmpc), the plume is roughly 130 by 20 kiloparsecs in extent,
with an integrated absolute R-band magnitude of -19.4.  To estimate
the colours, we restricted attention to an area of $80\arcsec \times
40\arcsec$ where the surface brightness is highest, near the eastern
end.  The removal of embedded features was done in a consistent way in
all colours.  These inclusions may in fact be part of the tidally
disrupted material, but it is impossible to determine this with
certainty from the low resolution Schmidt images; our conservative
approach was to measure only the smooth underlying light belonging to
the plume.  While the background can be determined to better than
$1\%$, this contributes an uncertainty of 0.1 to 0.2 magnitudes (10 to
20\%) in the colours because of the faintness of the feature.  The
plume's colours are B-V = 0.93, V-R = 0.57, and V-I = 1.17, consistent
with those of a normal elliptical galaxy.  For comparison, the normal
elliptical NGC3377$^{24}$ has B-V = 0.92, V-R = 0.58, V-I = 1.20 and
an absolute R magnitude$^{25}$ of -20.8.  These colours indicate little
if any recent star formation associated with the tidal event.
This constraint is not strong, however, because the colours have been
determined from a restricted area of the feature and the removed,
apparent foreground, objects may be star formation
activity in the plume.
}
\end{figure}

\begin{figure}
\plotfiddle{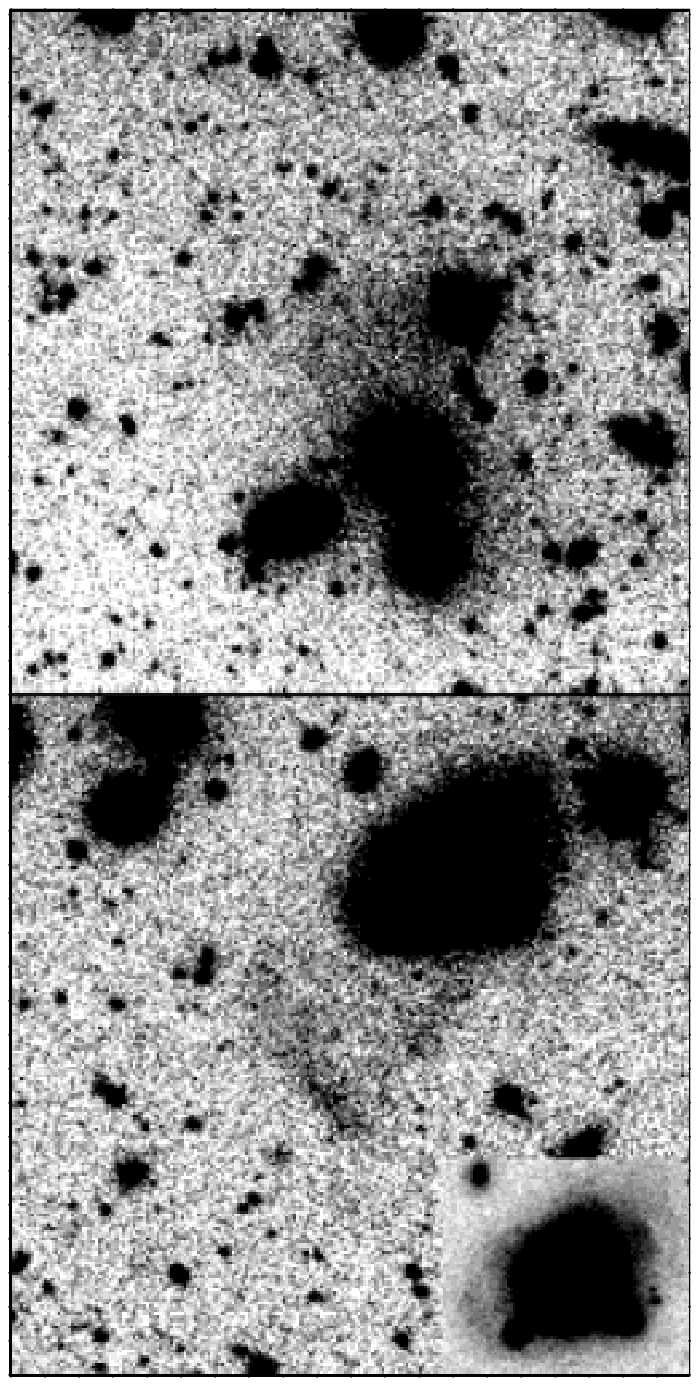}{5.75in}{0}{100}{100}{-160}{-40}
\end{figure}
\begin{figure}
\caption{LSB Objects no.~2 (upper) and no.~3 after removing foreground
stars and galaxies.  Each image is 400\arcsec\ square.  The colours and
absolute magnitude of no.~2 are B-V=0.90, V-R=0.43, V-I=1.06,
M$_R$=-18.8 and no.~3 are B-V=0.93, V-R=0.59, V-I=0.89, M$_R$=-18.3.
Like the plume feature, these are similar to a normal elliptical
galaxy and indicate little if any recent star formation, 
though here too, the removed foreground objects may in fact be part of
the tidal material.
The bright object immediately to the northwest of feature no.~2 is a
Galactic star.  The two objects immediately south of it are IC3959 and
IC3963; their low velocity difference of 240 km s$^{-1}$ may allow the strong
tidal interaction which gives rise to the LSB material.  The bright
object to the northwest of feature no.~3 is NGC4911, a face-on spiral.
The inset in the lower right is a lower contrast display of NGC4911
showing the low surface brightness material encircling its disk and
linking it to LSB Object no.~3.  NGC4911 has a close S0 companion which
is partly visible in the inset as a protrusion from the lower right of
the disk.  Their velocity difference is only 300 km/s, small enough to
permit strong tidal interactions.  }
\end{figure}



\begin{references}

\reference{ZW} 1. Zwicky, F. {\it Morphological Astronomy} p.\ 48
(Springer-Verlag, Berlin, 1957).

\reference{Biv} 2. Biviano, A. in {\it A New Vision of an Old
Cluster: Untangling Coma Berenices} (eds Mazure, A., Casoli F., Durret
F., \& Gerbal D) 1-8 (World Scientific Publishing, Singapore, 1998).
\reference{Mer84} 3. Merritt, D. Relaxation and tidal stripping in rich
clusters of galaxies. II - evolution of the luminosity distribution.
{\it Astrophys. J.} {\bf 276}, 26-37 (1984).

\reference{TM} 4. Trentham, N., \& Mobasher, B. The discovery of a giant
debris arc in the Coma cluster.  {\it Mon. Not. R. Astron. Soc.} {\bf
293}, 53-59 (1998)

\reference{WS} 5. Welch, G.A., \& Sastry, G.N. Photographic detection
of ``intergalactic'' matter in the Coma cluster.  {\it Astrophys. J.}
{\bf 169}, L3-L5 (1971).

\reference{SH97} 6. Secker, J., \& Harris, W.E. Dwarf galaxies in the
Coma cluster I. detection, measurement and classification techniques
{\it Publ. Astron. Soc. Pac.} {\bf 109}, 1364-1376 (1997).

\reference{BT} 7. Binney, J., Tremaine, S. {\em Galactic
Dynamics} (Princeton Univ.\ Press, 1987).

\reference{FW} 8. Fitchett, M.J., \& Webster, R.L. Substructure in the
Coma cluster. {\it Astrophys. J.}  {\bf 317}, 653-667 (1987).

\reference{Mel88} 9. Mellier, Y., Mathez, G., Mazure, A., Chauvineau, B.,
\& Proust, D. Subclustering and evolution of the Coma
cluster. {\em Astron.\ Astrophys.} {\bf 199}, 67-72 (1988).

\reference{WBH} 10. White, S.D.M., Briel, U.G., \& Henry, J.P.  X-ray
archaeology in the Coma cluster. {\it Mon. Not. R. Astron. Soc.}
{\bf 261}, L8-L12 (1993).

\reference{VFJ} 11. Vikhlinin, A., Forman, W., \& Jones, C.  Another
collision for the Coma cluster. {\it Astrophys. J.} {\bf 474}, L7-L10
(1997).

\reference{CD} 12. Colless, M. \& Dunn, A.M. Structure and dynamics of the
Coma cluster. {\it Astrophys. J.} {\bf 458}, 435-454 (1996).

\reference{CG} 13. Conselice, C.J., \& Gallagher, J.S. Galaxy
aggregates in the Coma cluster.  {\it Mon. Not. R. Astron. Soc.}
{\bf 297}, L34-L38 (1998).


\reference{Mih96} 14. Dubinski, J., Mihos, J.C., \& Hernquist,
L.  Using tidal tails to probe dark matter halos. {\it Astrophys. J.}
{\bf 462}, 576-593 (1996).

\reference{M96} 15. Moore, B., Katz, N., Lake, G., Dressler, A., \&
Oemler, A., Jr.  Galaxy harassment and the evolution of clusters of
galaxies. {\it Nature} {\bf 379}, 613-616 (1996).

\reference{NED} 16. {\it NASA/IPAC Extragalactic Database},\newline
nedwww.ipac.caltech.edu (1998).

\reference{TG} 17. Thompson, L.A., \& Gregory, S.A. Dwarf galaxies in the
Coma cluster. {\it Astron. J.} {\bf 106}, 2197-2212 (1993).

\reference{LC97} 18. Lopez-Cruz, O., Yee, H.K.C., Brown, J.P., Jones,
C., \& Forman, W.  Are luminous cD halos formed by the disruption of
dwarf galaxies? {\it Astrophys. J.} {\bf 475}, L97-L101 (1997).

\reference{T72} 19. Tinsley, B.M. A First Approximation to the Effect
of Evolution on $q_0$. {\it Astrophys. J.}
{\bf 173}, L93-L97 (1972).

\reference{OT} 20. Ostriker, J.P., \& Tremaine, S.D.  Another
evolutionary correction to the luminosity of giant galaxies. 
{\it Astrophys. J.} {\bf 202}, L113-L116 (1975).

\reference{TW} 21. Theuns, T. \& Warren, S.J. Intergalactic stars in the
Fornax cluster. {\it Mon. Not. R. Astron. Soc.} {\bf 284}, L11-L15 (1997).

\reference{FTvH} 22. Ferguson, H.C., Tanvir, N.R., \& von Hippel, T.
Detection of intergalactic red-giant-branch stars in the Virgo
cluster.  {\it Nature} {\bf 391}, 461-463 (1998).

\reference{CJFB} 23. Ciardullo, R., Jacoby, G.H., Feldmeier, J.J., \&
Bartlett, R.E.  The planetary nebula luminosity function of M87 and
the intracluster stars of Virgo.  {\it Astrophys. J.} {\bf 492}, 62-73 (1998).

\reference{W95} 24. West, M., Cote, P., Jones, C., Forman, W., \& Marzke,
R.  Intracluster globular clusters. {\it Astrophys. J.}
{\bf 453}, L77-L80 (1995).

\reference{G1} 25. Gregg, M.D. Differential population synthesis of S0
galaxies. I - the data.  {\it Astrophys. J. Suppl.} {\bf 69}, 217-232
(1989).

\reference{RSA} 26. Sandage, A., Tammann, G.A. {\em A revised
Shapley-Ames catalog of bright galaxies} (Carnegie Inst.\ of
Washington, 1987).

\end{references}
\end{document}